\documentclass[twoside]{article}
\usepackage{amsmath}
\usepackage{amssymb}
\usepackage{graphicx}
\usepackage{amsmath}
\usepackage{amsfonts}
\usepackage{euscript}
\input epsf
\usepackage{amsthm}
\def\enddemo{\qed \endtrivlist}
\expandafter\let\csname enddemo*\endcsname=\enddemo

\def\qedsymbol{\ifmmode\bgroup\else$\bgroup\aftergroup$\fi
\vcenter{\hrule\hbox{\vrule
height.6em\kern.6em\vrule}\hrule}\egroup}
\def\qed{\ifmmode\else\unskip\nobreak\fi\quad\qedsymbol}

\newtheorem{thm}{Theorem}[section]

\newtheorem{lem}{Lemma}[section]

\newtheorem{exm}{Example}[section]

\newtheorem{dfn}{Definition}[section]
\newtheorem{alg}{Algorithm}[section]

\newcommand{\ra}[1]{\textrm{rank}({#1})} % rank
\newcommand{\oo}[1]{{\breve{#1}}}
\newcommand{\ov}[1]{{\widetilde{#1}}}
\newcommand{\e}{\mathcal{E}}
\newcommand{\Eff}{\mathrm{\bf Eff}}
\newcommand{\Ind}{\mathrm{Ind}}
\newcommand{\Sp}{\mathrm{\bf Sp}}
\newcommand{\Ef}{\mathrm{\bf Ef}}
\newcommand{\ef}{\mathrm{\bf ef}}
\newcommand{\col}{\mathrm{\bf col}}
\newcommand{\row}{\mathrm{\bf row}}
\newcommand{\conjugate}[1]{\overline{#1}}
\renewcommand{\O}{\mathcal{O}}
\renewcommand{\widehat}[1]{#1}

 \font\ssr=cmss8 
\font\sst=cmtt8   \font\ssi=cmti8
 \font\ssr=cmss8 
\font\sst=cmtt8   \font\ssi=cmti8

\title{\bf Effective partitioning method \\ for computing \\ weighted Moore-Penrose inverse}
\begin{footnotesize}
\author{\frenchspacing
\bf Marko D. Petkovi\' c\footnote{Corresponding author}, Predrag S. Stanimirovi\' c, Milan B. Tasi\' c\\
{\ssi University of Ni\v{s}, Department of Mathematics, Faculty of Science,}\\
{\ssi Vi\v segradska 33, 18000 Ni\v s, Serbia} \\
{\ssi E-mail:} {\sst dexter\_of\_nis@neobee.net},\ \ {\sst pecko@pmf.ni.ac.yu},\ \ {\sst milan12t@ptt.yu} \\
}
\end{footnotesize}
\date{}
\begin{document}

\maketitle

\begin{abstract}
We introduce a method and algorithm for computing the weighted Moore-Penrose
inverse of multiple-variable polynomial matrix and the related algorithm which is appropriated for sparse polynomial matrices.
These methods and algorithms are generalizations of algorithms developed in \cite{TSP}
to multiple variable rational and polynomial matrices and improvements of these algorithms on sparse matrices.
Also, these methods are generalizations of the partitioning method for computing the
Moore-Penrose inverse of rational and polynomial matrices introduced in \cite{Stanimirovic2} and \cite{Stanimirovic3}
to the case of weighted Moore-Penrose inverse.
Algorithms are implemented in the symbolic computational package {\ssr MATHEMATICA\/}.

\frenchspacing \itemsep=-1pt
\begin{description}
\item[] AMS Subj. Class.: 15A09, 68Q40.
\item[] Key words: Weighted Moore-Penrose inverse; rational matrices, polynomial matrices; sparse matrices; symbolic computation.
\end{description}
\end{abstract}

\section{Introduction}
Let $\mathbf C^{m\times n}$ be the set of $m \times n $ complex matrices, and ${\mathbf C}^{m\times n}_r$
is the set of $m \times n $ complex matrices of rank $r$:
${\mathbf C}^{m\times n}_r\!=\!\{X\in {\mathbf C}^{m \times n}\, |\,\,\, \ra{X}\!=\!r\}$.
For any matrix $A\in\mathbf {C}^{m \times n}$ and
positive definite Hermitian matrices $M$ and $N$ of the order $m$ and $n$ respectively,
consider the following equations in $X$, where $*$ denotes conjugate and transpose:
$$\begin{array}{ll}
(1)\quad AXA=A & (2)\quad XAX\! =\! X \\
(3)\quad (MAX)^*=MAX & (4)\quad (NXA)^*\! =\! NXA.
\end{array}$$
The matrix $X$ satisfying these equations is called the weighted Moore-Penrose inverse of
$A$, and it is denoted by $X=A_{MN}^{\dagger}$. In the partial
case $M=I_m$, $N=I_n$, the matrix $X=A_{MN}^{\dagger}$ comes to
the Moore-Penrose inverse $A^{\dagger}$ of $A$.

\smallskip
As usual, $\mathbf C[s_1,\ldots,s_p]$ (resp. $\mathbf C(s_1,\ldots,s_p)$) denotes the
polynomials (resp. rational functions) with complex coefficients
in the variables $s_1,\ldots,s_p$. The matrices of format $m\!\times \!n$ with elements
in $\mathbf C[s_1,\ldots,s_p]$ (resp. $\mathbf C(s_1,\ldots,s_p)$) are denoted by $\mathbf
C[s_1,\ldots,s_p]^{m\times n}$ (resp $\mathbf C(s_1,\ldots,s_p)^{m\times n}$).
By $I$ it is denoted an appropriate identity matrix.

\smallskip
Computation of the Moore-Penrose inverse of one variable
polynomial and/or rational matrices, based on the
Leverrier-Faddeev algorithm, is investigated in
\cite{Bar},\cite{Fra},\cite{Jon},\cite{Kar},\cite{Pecko1},\cite{Tze}. Implementation of this
algorithm in the symbolic computational language {\ssr MAPLE\/},
is described in \cite{Jon}. Algorithm for computing the
Moore-Penrose inverse of two-variable rational and polynomial
matrix is introduced in \cite{Kar4}. A quicker and less
memory-expensive Effective algorithm for computing the
Moore-Penrose inverse of one-variable and two-variable polynomial
matrix, with respect to those introduced in \cite{Kar} and
\cite{Kar4}, is presented in \cite{Kar2}. This algorithm is efficient when elements of the input matrix are polynomials
with only few nonzero addends.

\smallskip
Papers \cite{Bu},\cite{Ji},\cite{Pecko1} deal with a computation of the {\it Drazin inverse}.
A generalization of these algorithms, introduced in
\cite{Pecko2}, generates the wide class of outer
inverses of a rational or polynomial matrix.

\smallskip
An interpolation algorithm for computing the Moore-Penrose inverse
of a given one-variable polynomial matrix, based on the
Leverrier-Faddeev method, is presented in \cite{Marko}. Algorithms
for computing the Moore-Penrose and the Drazin inverse of
one-variable polynomial matrices based on the
evaluation-interpolation technique and the Fast Fourier transform
are introduced in \cite{Kar3}. Corresponding algorithms for
two-variable polynomial matrices are introduced in \cite{Vol}.

\smallskip
In this paper we consider the set of rational and
polynomial matrices and various variants of the partitioning
method for computing generalized inverses. Grevile's {\it partitioning method\/} for numerical computation of generalized
inverses is introduced in \cite{Gre}. Two different proofs for
Greville's method were presented in \cite{Campbell}, \cite{Wang}.
A simple derivation of the Grevile's result has been given by
Udwadia and Kalaba \cite{Udwadia}. In \cite{Fan} Fan and Kalaba
used the approach of determination of the Moore-Penrose inverse of
matrices using dynamic programming and Belman's principle of optimality.
Wang in \cite{Wang1} generalizes Grevile's method to the weighted Moore-Penrose inverse.

\smallskip
Many numerical algorithms for computing the Moore-Penrose inverse
lack numerical stability.
The Greville's algorithm requires more operations and consequently it accumulates more rounding
errors (see for example \cite{Shi}). Moreover, it is well-known that the Moore-Penrose inverse
is not necessarily a continuous function of the elements of the
matrix. The existence of this discontinuity present further
problems in the pseudoinverse computation. It is therefore clear
that cumulative round off errors should totally eliminated, which is possible only by means of
the symbolic implementation. In the symbolic implementation variables are stored in the "exact"
form or can be left "unassigned" (without numerical values),
resulting in no loss of accuracy during the calculation \cite{Kar}.

\smallskip
An algorithm for computing the Moore-Penrose inverse of
one-variable polynomial and/or rational matrices, based on the Grevile's partitioning algorithm, is introduced in
\cite{Stanimirovic2}. An extension of results from \cite{Stanimirovic2} to the set of two-variable rational and
polynomial matrices is introduced in the paper \cite{Stanimirovic3}.
In our recent paper \cite{TSP} we propose an algorithm for computing the weighted Moore-Penrose of one-variable
rational and polynomial matrix. In this work we generalized the results from \cite{TSP} in the following two ways:
\begin{itemize}
\item[-] extends algorithms from \cite{TSP} to the set of multi-variable rational and polynomial matrices with complex coefficients,
\item[-] make algorithms from \cite{TSP} more effective on sparse matrices with a relatively small number of nonzero elements.
\end{itemize}
The structure of the paper is as follows. In the second section we
extend the algorithm for computing the weighted Moore-Penrose from
\cite{Wang1} to the set of multiple-variable rational matrices with complex coefficients.
Main results are given in the third and the fourth section.
In Section 3 we adapt previous algorithm to the set of polynomial matrices.
In the fourth section we consider two effective structures which exploit only nonzero addends in
polynomial matrices and improve previous results on the set of sparse matrices.
In the last section we presented an illustrative example and compared various algorithms.

\section{Weighted Moore-Penrose inverse for multi-\\variable rational matrices} \setcounter{equation}{0}

Let $A(s_1,\ldots,s_p)$ be complex rational matrix. For the sake of
simplicity, we will introduce new variables $s_{2p+1-i}=\overline{s_i}$.
Also we will denote the vector of all variables $s_1,\ldots,s_{2p}$ by $S=(s_1,\ldots,s_{2p})$ and
further we will denote $A(s_1,\ldots,s_p)$ as $A(S)$.

\smallskip
By $\widehat{A}_i(S)$ we denote the submatrix of $A(S)$ consisting of its first $i$ columns, and
by $a_i(S)$ is denoted the $i$-th column of $A(S)$:
\begin{equation}\label{dva1}
\widehat{A}_i(S)=\left[ \widehat{A}_{i-1}(S)\ |\
a_i(S)\right],i=2,\ldots ,n,\quad \widehat{A}_1(S)=a_1(S)
\end{equation}
We will consider positive definite Hermitian matrices $M(S)\in\mathbf {C}(S)^{m \times m}$
and $N(S)\in\mathbf {C}(S)^{n \times n}$. The leading principal
submatrix $N_i(S)\in\mathbf {C}(S)^{i\times i}$ of $N(S)$ is partitioned as
\begin{equation}\label{dva2}
N_i(S)=\left[
\begin{array}{ll}
N_{i-1}(S) & l_i(S) \\
l_i^*(S) & n_{ii}(S) \\
\end{array}
\right] ,\ i=2,\ldots ,n,
\end{equation}
where  $l_i(S)\in\mathbf{C}(S)^{(i-1) \times 1}$ and $n_{ii}(S)$ is the complex polynomial. By $N_1(S)$ we denote the
polynomial $n_{11}(S)$.

\smallskip
In the following lemma we generalize the representations of the
weighted Moore-Penrose inverse from \cite{Wang},\cite{TSP} to the set of
rational matrices of multiple complex variables $\mathbf{C}(S)^{m \times n}$.

\smallskip
For the sake of simplicity, by $X_i(S)$ we denote the weighted
Moore-Penrose inverse corresponding to $M(S)$ and submatrices
$\widehat{A}_i(S)$, $N_i(S)$:
$X_i(S)=\widehat{A}_i(S)^\dagger_{MN_i}$, for each $i=2,\ldots ,n$. Similarly $X_1(S)=a_{1}(S)^{\dagger}_{M,N_1}$

\begin{lem}\label{lem1}
Let $A(S)\!\!\in\!\!\mathbf {C}(S)^{m \times n}$, assume that
$M(S)\!\!\in\!\!\mathbf {C}(S)^{m \times m}$, $N(S)\!\!\in\!\!\mathbf {C}(S)^{n \times n}$
are positive definite Hermitian matrices, and let
$\widehat{A}_i(S)$ be the submatrix of $A(S)$ consisting of its
first $i$ columns, as it is defined in $(\ref{dva1})$. Assume that the leading principal submatrix
$N_i(S)\in\mathbf {C}(S)^{i \times i}$ is partitioned as in $(\ref{dva2})$.
Then the matrices $X_i(S)$ can be computed in this way:

\begin{equation}\label{dva3}
X_1(S)= \left\{ \begin{array}{lc}
\left( a_{1}^*(S) M(S) a_{1}(S)\right) ^{-1} a_{1}^*(S) M(S), & a_{1}(S)\ne 0,\\
a_1^*(S), & a_{1}(S)=0,
\end{array}
\right.
\end{equation}
\begin{eqnarray}\label{dva4}
&& X_{i}(S)\!\!=\!\!
\left[ \begin{array}{c}
 X_{i-1}(S)\!-\!\left( d_i(S)\!+\!(I\!-\!X_{i-1}(S)A_{i-1}(S)\right) N_{i-1}^{-1}(S)l_i(S))b_i^*(S) \\
b_i^*(S) \end{array}\right], \nonumber\\
&& i=2,\ldots ,n,
\end{eqnarray}
where the vectors $d_i(S)$, $c_i(S)$ and $b_i^*(S)$ are defined by
\begin{eqnarray}
d_i(S) &=& X_{i-1}(S)a_i(S)\label{dva5} \\
c_i(S) &=& a_i(S)-A_{i-1}(S)d_i(S) = \left( I-A_{i-1}(S)X_{i-1}(S)\right) a_i(S)\label{dva6}
\end{eqnarray}
\begin{equation}\label{dva7}
b_i^*(S)=\left\{
\begin{array}{ll}
\left( c^*_i(S) M(S) c_i(S)\right) ^{-1}c_i^*(S) M(S), & c_i(S)\ne 0 \\\\
\delta_i^{-1}(S)\left( d_i^*(S) N_{i-1}(S)-l_i(S)^*\right) X_{i-1}(S), & c_i(S)=0,
\end{array}
\right.
\end{equation}
and where in $b_i^*(S)$ is
\begin{eqnarray}\label{dva8}
\delta_i(S)&=&n_{ii}(S)+d^*_i(S) N_{i-1}(S) d_i(S)-\left( d^*_i(S) l_i(S)+l_i^*(S)d_i(S)\right) \nonumber\\
&&\ -l_i^*(S)\left( I-X_{i-1}(S) A_{i-1}(S)\right) N_{i-1}^{-1}(S)l_i(S).
\end{eqnarray}
\end{lem}

Also in \cite{Wang} authors used a block representation of the inverse $N^{-1}_i(S)$,
which we also generalized to the set of rational matrices.

\begin{lem}\label{lem2}
Let $N_i(S) $ be the partitioned matrix defined in $(\ref{dva2})$.
Assume that $N_i(S)$ and $N_{i-1}(S)$ are both nonsingular. Then
\begin{eqnarray} \label{dva9}
N_i^{-1}(S)\!=\!\left\{\begin{array}{ll}
\left[\begin{array}{ll}
N_{i-1}(S) & l_i(S) \\
l_i^*(S) & n_{ii}(S)
\end{array}\right]^{-1}=
\left[
\begin{array}{ll}
E_{i-1}(S) & f_i(S) \\
f_i^*(S) & h_{ii}(S)
\end{array}
\right], & i\!=\!2,\ldots,n,\\{}\\
n_{11}^{-1}(S), & i=1,
\end{array} \right.
\end{eqnarray}
where
\begin{eqnarray}
h_{ii}(S)\!&=&\!\left( n_{ii}(S)-l_i^*(S)N_{i-1}^{-1}(S)l_i(S)\right) ^{-1} \label{dva10}\\
f_i(S)\!&=\!&-h_{ii}(S)N_{i-1}^{-1}(S)l_i(S)\label{dva11}\\
E_{i-1}(S)\!&=&\!N_{i-1}^{-1}(S)+h_{ii}^{-1}(S)f_i(S)f_i^*(S).\label{dva12}
\end{eqnarray}
\end{lem}

In view of Lemma {\ref {lem1}} and Lemma {\ref {lem2}}, respectively, we present the following algorithms for computing
the weighted Moore-Penrose inverse and the inverse matrix $N_{i}^{-1}(S)\in \mathbf C(S)^{i\times i}$.
These algorithms are generalizations of corresponding algorithms from \cite{TSP} to the set of multiple-variable
rational matrices with complex coefficients.

\begin{alg}\label{alg21}

Input: $A(S)\!\in \!\mathbf C(S)^{m\times n}$ and positive definite
matrices $M(S)\!\in \!\mathbf C(S)^{m\times m}$ and $N(S)\!\in \!\mathbf C(S)^{n\times n}$.

\smallskip
\leftskip 0.5cm {\it Step 1.\/} Initial value: Compute
$X_1(S)=a_1(S)^{\dagger}$ defined in $(\ref{dva3})$.

\smallskip
{\it Step 2.\/} Recursive step: For each $i=2,\ldots ,n$ compute
$X_i(S)$ performing the following four steps:

\leftskip 0.8cm

\smallskip
{\it Step 2.1.\/} Compute $d_i(S)$ using $(\ref{dva5})$.

\smallskip
{\it Step 2.2.\/} Compute $c_i(S)$ using $(\ref{dva6})$.

\smallskip
{\it Step 2.3.\/} Compute $b_i^*(S)$ by means of $(\ref{dva7})$ and $(\ref{dva8})$.

\smallskip
{\it Step 2.4.\/} Applying $(\ref{dva4})$ compute $X_i(S)$.

\leftskip 0.5cm

\smallskip
{\it Step 3.\/} The stopping criterion: $i=n$. Return $X_n(S)$.

\leftskip 0cm

\end{alg}

\begin{alg}\label{alg22}
Let $N_i(S)\!=\!\left[\begin{array}{ll}
N_{i-1}(S) & l_{i}(S) \\
l_{i}^*(S) & n_{ii}(S)\end{array}\right]$
be the leading principal submatrix of positive definite matrix $N\in \mathbf C(S)^{n\times n}$.
Then the inverse matrix $N^{-1}(S)$ can be computed as follows:

\smallskip
\leftskip 0.5cm {\it Step 1.\/} Initial values:
$N_{1}^{-1}(S)=n_{11}^{-1}(S)$.

\leftskip 0.5cm

\smallskip
{\it Step 2.\/} Recursive step: For $i=2,\ldots ,n$ perform the
following steps:

\leftskip 0.8cm
\smallskip
{\it Step 2.1.\/} Compute $h_{ii}(S)$ using $(\ref{dva10})$.

\smallskip
{\it Step 2.2.\/} Compute $f_i(S)$ using $(\ref{dva11})$.

\smallskip
{\it Step 2.3.\/} Compute $E_{i-1}(S)$ using $(\ref{dva12})$.

\smallskip
{\it Step 2.4.\/} Compute $N_i^{-1}(S)$ using $(\ref{dva9})$.

\leftskip 0.5cm
\smallskip
{\it Step 3.\/} For $i=n$ return the inverse matrix $N^{-1}(S)=N_n^{-1}(S)$.

\leftskip 0cm

\end{alg}

We used {\ssr MATHEMATICA} function {\it Together} in order to enable
simplifications of rational expressions (this function joins
rational addends together and cancels common multipliers in numerator and denominator).

\section{Weighted Moore-Penrose inverse for multi-\\variable polynomial matrices}\setcounter{equation}{0}

Now suppose that $A(S) \in \mathbf{C}[S]^{m \times n}$ is
multi-variable polynomial matrix. We can represent it in the following polynomial form:
\begin{equation}\label{tri1}
A(S)=\sum_{i_1=0}^{d_1} \cdots \sum_{i_{2p}=0}^{d_{2p}} A_{i_1,\ldots,i_{2p}}s_1^{i_1} \cdots s_{2p}^{i_{2p}}=\sum\limits_{I=0}^Q A_IS^I,
\end{equation}
where $I\!=\!(i_1,\ldots,i_{2p})$, $A_I\!=\!A_{i_1,\ldots,i_{2p}}$ are constant $m \times n$ matrices,
 $S^I\!=\!s_{1}^{i_1}s_{2}^{i_2} \cdots s_{2p}^{i_{2p}}$,
$Q=(d_1,\ldots,d_{2p})={\mathrm deg} A(S)$. Here $d_i$ is the degree of the matrix polynomial with respect to the variable $s_i$ in $A(S)$.

\smallskip
If by $\overline{J}$ we denote $\overline{J}=(j_{2p},\ldots,j_1)$, where $J=(j_1,\ldots,j_{2p})$ then it can be
easily checked that holds
$A^*(S)=\sum\limits_{J=0}^{\overline{Q}}A^*_J S^J$.

\smallskip
An application of Algorithm 2.1 to the multiple-variable polynomial matrix $A(S)$
gives the following result.

\begin{thm}
Let us consider $A(S)\!\in \!{\mathbf C}[S]^{m\times n}$ of the form $(\ref{tri1})$ and positive definite Hermitian matrices
$M(S)\!\in\!\mathbf {C}(S)^{m \times m}$ and $N(S)\in\mathbf {C}(S)^{n \times n}$.
Assume that the leading principal submatrix $N_i(S)\in\mathbf {C}(S)^{i\times i}$ of $N(S)$ is partitioned as in $(\ref{dva2})$.
Then the weighted Moore-Penrose inverse $A_{MN_i}^\dagger(S) \in {\mathbf C}^{i\times m}[S]$
corresponding to the first $i$ columns in $A(S)$ is of the form
\begin{equation}
X_i(S)=A_{MN_i}^\dagger(S)=\frac{Z_i(S)}{Y_i(S)},\ i=1,\ldots ,n,
\end{equation}
where $Z_i(S) \in \mathbf{C}^{m \times i}[S]$ and $Y_i(S) \in \mathbf{C}[S]$,
can be computed from $Z_{i-1}(S)$, $Y_{i-1}(S)$, $A_{i-1}(S)$ and $a_i(S)$ using exact recurrence relations.
\end{thm}
\begin{demo}
We will prove theorem by the induction. In the case $i=1$ exact relations for
$Z_1(S)$ and $Y_1(S)$ can be derived from $(\ref{dva3})$:
$$
\aligned
a_1(S)=A_1(S)&=0 \Rightarrow Z_1(S)=0, \quad Y_1(S)=1 \\
a_1(S)=A_1(S)&\neq 0 \Rightarrow Z_1(S)=a^*_1(S)M(S), \quad Y_1(S)=a^*_1(S)M(S)a_1(S)
\endaligned
$$
Consider now the inductive step. From the inductive hypothesis we can write $X_{i-1}(S)=\frac{Z_{i-1}(S)}{Y_{i-1}(S)}$.
Then $X_i(S)$ can be computed by using Step 2 of algorithm 2.1. From steps 2.1 and 2.2 we have:
$$
\aligned
d_i(S)&=X_{i-1}(S)a_i(S)=\frac{Z_{i-1}(S)a_i(S)}{Y_{i-1}(S)}=\frac{D_i(S)}{Y_{i-1}(S)}\\
c_i(S)&=a_i(S)-A_{i-1}(S)d_i(S)=\frac{a_i(S)Y_{i-1}(S)-A_{i-1}(S)D_i(S)}{Y_{i-1}(S)}=\frac{C_i(S)}{Y_{i-1}(S)}.
\endaligned
$$
If $C_i(S)\neq 0$, according to the Step 2.3 of Algorithm 2.1 we have:
$$
b^*_i(S)=\frac{\frac{C^*_i(S)}{Y^*_{i-1}(S)}M(S)}{\frac{C^*_i(S)}{Y^*_{i-1}(S)}M(S)\frac{C_i(S)}{Y_{i-1}(S)}}=
\frac{Y_{i-1}(S)C^*_i(S)M(S)}{C^*_i(S)M(S)C_i(S)}=\frac{V_i(S)}{W_i(S)}
$$
Otherwise, we need first to evaluate the expression $\delta_i(S)$. From (\ref{dva8}) we obtain:
\begin{eqnarray}
\delta_i(S)\!\!\!\!&=&\!\!\!\!n_{ii}(S)+\frac{D^*_i(S)}{Y^*_{i-1}(S)} N_{i-1}(S) \frac{D_i(S)}{Y_{i-1}(S)}\nonumber \\
&&\!\!\!-\!\!\left(\! \frac{D^*_i(S)}{Y^*_{i\!-\!1}(S)}l_i(S)\!+\!l_i^*(S)\frac{D_i(S)}{Y_{i\!-\!1}(S)}\!\right)
\!-\!l_i^*(S) \frac{\phi_i(S)}{\psi_i(S)}. \label{deltapoly}
\end{eqnarray}
Here we used the inductive hypothesis together with
temporary polynomial matrix $\phi_i(S) \in \mathbf{C}[S]^{(i-1) \times 1}$ and polynomial $\psi_i(S)$ are defined by:
\begin{equation}
\aligned
&\left( I-X_{i-1}(S) A_{i-1}(S)\right) N_{i-1}^{-1}(S)l_i(S)\\
&=\frac{Y_{i-1}(S)I-Z_{i-1}(S)A_{i-1}(S)}{Y_{i-1}(S)}\cdot
\frac{\ov{N}_{i-1}(S)}{\oo{N}_{i-1}(S)}\cdot l_i(S)\\
&=\frac{Y_{i-1}(S)\ov{N}_{i-1}(S)l_i(S)-Z_{i-1}(S)A_{i-1}(S)\ov{N}_{i-1}(S)l_i(S)}{Y_{i-1}(S)\oo{N}_{i-1}(S)}
=\frac{\phi_i(S)}{\psi_i(S)}.
\endaligned \label{fipsi}
\end{equation}
Also, we use $N_{i-1}^{-1}(S)=\frac{\ov{N}_i(S)}{\oo{N}_i(S)}$, where $\ov{N}_i(S)\in \mathbf{C}[S]^{(i-1) \times (i-1)}$ and
$\oo{N}_i(S)\in \mathbf{C}[S]$ are defined in the next theorem.
By collecting addends under the same denominator in $(\ref{deltapoly})$ we can write $\delta_i(S)$ in the form:
$$
\delta_i(S)=\frac{\ov{\Delta}_i(S)}{\oo{\Delta}_i(S)}
$$
where:
\begin{eqnarray*}
\ov{\Delta}_i(S)\!\!\!\!&=&\!\!\!n_{ii}(S)\oo{N}_{i-1}(S)Y^*_{i-1}(S)Y_{i-1}(S)\!+\!\oo{N}_{i-1}(S)D^*_i(S)N_{i-1}(S)D_i(S)\\
&&\!\!\!-\!\!\left(Y_{i\!-\!1}(S)D^*_i(S)l_i(S)\!+\!Y^*_{i-1}(S)D_i(S)l^*_i(S) \right)\oo{N}_{i-1}(S)\!-\!l^*_i(S)\phi_i(S)Y^*_{i-1}(S)\\
\oo{\Delta}_i(S)\!\!&=&\!\!Y^*_{i-1}(S)Y_{i-1}(S)\oo{N}_{i-1}(S).
\end{eqnarray*}
Now we apply Step 2.3 in the case $C_i(S)=0$ and evaluate $b^*_i(S)$:
$$
\aligned
b^*_i(S)&=\frac{\ov{\Delta}_i(S)}{\oo{\Delta}_i(S)}\left( \frac{D^*_i(S)}{Y^*_{i-1}(S)} N_{i-1}(S)-l_i(S)^*\right)
\frac{Z_{i-1}(S)}{Y_{i-1}(S)}\\
&=\frac{\ov{N}_{i-1}(S)\left( D^*_i(S)N_{i-1}(S)- Y^*_{i-1}(S)l^*_i(S) \right)Z_{i-1}(S)}{\oo{\Delta}_i(S)}=\frac{V_i(S)}{W_i(S)}.
\endaligned
$$
Let us rewrite now expression (\ref{dva4}) in following way:
\begin{eqnarray*}
X_i(S)\!\!\!&=&\!\!\!\bmatrix \frac{Z_{i-1}(S)}{Y_{i-1}(S)}-\left(\frac{D_i(S)}{Y_{i-1}(S)}+\frac{\phi_i(S)}{\psi_i(S)}\right)
\frac{V_i(S)}{W_i(S)} \\
{}\\
\frac{V_i(S)}{W_i(S)} \endbmatrix\\
&&{}\\
&=&\!\!\!\frac{1}{W_i(S)\psi_i(S)}\bmatrix W_i(S)\oo{N}_{i-1}(S)Z_{i-1}(S)\!-\!\left(D_i(S)\oo{N}_{i-1}(S)+\phi_i(S)\right)V_i(S) \\
\oo{N}_{i-1}(S)Y_{i-1}(S)V_i(S) \endbmatrix .
\end{eqnarray*}
>From the last expression we obviously have that holds:
\begin{eqnarray*}
Z_i\!\!\!&=&\!\!\!\bmatrix W_i(S)\oo{N}_{i-1}(S)Z_{i-1}(S)-\left(D_i(S)\oo{N}_{i-1}(S)+\phi_i(S)\right)V_i(S) \\
\oo{N}_{i-1}(S)Y_{i-1}(S)V_i(S) \endbmatrix \!= \! \bmatrix \Theta_i(S) \\ \Psi_i(S) \endbmatrix\\
Y_i\!\!\!&=&\!\!\!W_i(S)\psi_i(S).
\end{eqnarray*}
This completes the proof of the theorem.
\end{demo}

\begin{thm}
Let the leading principal submatrix $N_i(S)$ of the positive definite matrix $N(S) \in \mathbf{C}[s]^{n \times n}$
is partitioned as in $(\ref{dva2})$. Then the inverse $N^{-1}_i(S)$ is of the form:
$$
N^{-1}_i(S)=\frac{\ov{N}_i(S)}{\oo{N}_i(S)}=\frac1{\oo{N}_i(S)}
\bmatrix
E_{i-1}(S) & F_i(S) \\
F_i^*(S) & H_{ii}(S)
\endbmatrix
$$
where $E_{i-1}(S) \in \mathbf{C}^{(i-1) \times (i-1)}$, $F^*_i(S) \in \mathbf{C}^{(i-1) \times 1}$ and scalar
$H_{ii}(S) \in \mathbf{C}[s]$ can be
computed from $N_{i-1}(S)$, $l(S)$, $n_{ii}(S)$, $\ov{N}_{i-1}(S)$ and $\oo{N}_{i-1}(S)$ using exact recurrence relations.
\end{thm}

\begin{demo} As in the proof of the previous theorem we will use induction and lemma 2.2 (algorithm 2.2). The case
$i=1$ is again trivial and we have:
$$\ov{N}_1(S)=1, \quad \oo{N}_1(S)=n_{11}(S)$$
Let us consider now the inductive step and suppose that $N^{-1}_{i-1}(S)=\frac{\ov{N}_{i-1}(S)}{\oo{N}_{i-1}(S)}$. From the relation
(\ref{dva10}) we have:
\begin{eqnarray}\label{Hii}
\frac1{H_{ii}(S)}\!\!&=&\!\!n_{ii}(S)-l^*_i(S)\frac{\ov{N}_{i-1}(S)}{\oo{N}_{i-1}(S)}l_i(S)\\
&= &\!\!\frac{\oo{N}_{i-1}(S)n_{ii}(S)-l^*_i(S)\ov{N}_{i-1}(S)l_i(S)}{\oo{N}_{i-1}(S)}=\frac{\oo{H_i}(S)}{\oo{N}_{i-1}(S)}.\nonumber
\end{eqnarray}

Therefore, we can write $H_{ii}(S)=\frac{\oo{N}_{i-1}(S)}{\oo{H_i}(S)}$. Using the relation (\ref{dva11}) we can represent $f_i(S)$
in following way:
$$f_i(S)=-\frac{\ov{N}_{i-1}(S)}{\oo{H_i}(S)}\cdot l^*_i(S) \cdot \frac{\ov{N}_{i-1}(S)}{\oo{N}_{i-1}(S)}=
-\frac{l^*_i(S)\ov{N}_{i-1}(S)}{\oo{H}_i(S)}=\frac{\ov{F}_i(S)}{\oo{H}_i(S)}.$$

Furthermore using the fact that $\ov{N}_{i-1}(S)$ is symmetric and positive definite, we can conclude that
$\ov{F}^*_i(S)=\ov{N}_{i-1}(S)l_i(S)$
which further implies that:
$$f^*_i(S)=\frac{\ov{F}^*_i(S)}{\oo{H}^*_i(S)}=\frac{\ov{N}_{i-1}(S)l_i(S)}{\oo{H}_i(S)}.$$
We also used that $\oo{H}_i(S)=\oo{H}^*_i(S)$ which can be easily proven from (\ref{Hii}). From (\ref{dva12}) we can conclude:
\begin{eqnarray*}
E_{i-1}\!\!&=&\!\!\frac{\ov{N}_{i-1}(S)}{\oo{N}_{i-1}(S)}+
\frac{\oo{H}_i(S)}{\oo{N}_{i-1}(S)}\frac{\ov{F}_i(S)}{\oo{H}_i(S)}\frac{\ov{F}^*_i(S)}{\oo{H}_i(S)}\\
&=&\!\!\frac{\ov{N}_{i-1}(S)-\ov{F}_i(S)\ov{F}^*_i(S)}{\oo{N}_{i-1}(S)\oo{H}_i(S)}
\!=\!\frac{\ov{E}_{i-1}(S)}{\oo{N}_{i-1}(S)\oo{H}_i(S)}.
\end{eqnarray*}
Finally, we can represent $N^{-1}_i(S)$ in the following matrix form:
\begin{eqnarray*}
N^{-1}_i(S)\!\!&=&\!\!\bmatrix
\frac{\ov{E}_{i-1}(S)}{\oo{N}_{i-1}(S)\oo{H}_i(S)} & \frac{\ov{F}_i(S)}{\oo{H}_i(S)} \\
{}& {}\\
\frac{\ov{F}^*_i(S)}{\oo{H}_i(S)} & \frac{\oo{N}_{i-1}(S)}{\oo{H}_i(S)}
\endbmatrix\\
&=&\!\!\frac1{\oo{H}_i(S)\oo{N}_{i-1}(S)}
\bmatrix
\ov{E}_{i-1}(S) & \oo{N}_{i-1}(S) \ov{F}_i(S) \\
\oo{N}_{i-1}(S)\ov{F}^*_i(S) & \oo{N}_{i-1}(S)^2
\endbmatrix=\frac{\ov{N}_i(S)}{\oo{N}_i(S)}.
\end{eqnarray*}
This completes proof of the theorem.
\end{demo}

Now it is easy to construct corresponding algorithms from the theorems 3.1 and 3.2.

\section{Effective method}\setcounter{equation}{0}

In practice we often work with polynomial matrices $A(S)$ with
a relatively small number of nonzero coefficients. In that case, previous algorithm is
not effective because of many operations are redundant.
To avoid this problem we will construct two appropriate sparse structures for the representation of the polynomial matrix $A(S)$
and corresponding effective algorithm for computing $A_{MN}^\dagger(S)$.
The first sparse representation is denoted by $\Eff$ and its improvement by $\Eff'$, while the second structure is denoted by $\Ef$.

\smallskip
The main idea in the first considered sparse structure is to exploit only non-zero coefficient matrices
$A_I=A_{i_1,\ldots,i_{2p}} \neq 0$
of the polynomial matrix $A(S)$ given in the form (\ref{tri1}).

\begin{dfn}
The {\it effective sparse structure} of the polynomial matrix $A(S)$, defined in $(\ref{tri1})$, is equal to:
\begin{equation}\label{cetiri1}
\Eff_A=\left\{(J,A_J)\,|\,A_J \neq 0,\ 0 \leq J \leq \deg A(S)\right\}.
\end{equation}
Also define {\it the index set} of this effective structure by:
\begin{equation}
\Ind_A=\left\{J\,|\,A_J \neq 0,\ 0 \leq J \leq \deg A(S)\right\}.
\end{equation}
Define operations $+$, $-$, $\cdot$ and $*$ on sparse structures as:
\begin{equation}
\aligned
\Eff_A+ \Eff_B&=\Eff_{A+B},\ \Eff_A- \Eff_B=\Eff_{A-B},\\
\Eff_A\cdot \Eff_B&=\Eff_{A\cdot B}, \quad \Eff^*_A=\Eff_{A^*}.
\endaligned \label{Eff}
\end{equation}
Denote by $e_A=|\Eff_A|=|\Ind_A|$ the {\it size} of the structure $\Eff_A$.
\end{dfn}

Obviously we have
$$A(S)\cdot B(S)=\sum\limits_{\begin{footnotesize}\begin{array}{rr} I \in \Ind_A\\ J \in \Ind_B \end{array} \end{footnotesize}}
  A_IB_JS^{I+J},$$
where
$$S^{I+J}=s_1^{i_1+j_1}\cdots s_{2p}^{i_{2p}+j_{2p}}. $$
If $C(S)=A(S)B(S)$ then the elements of
$\Eff_C$ are pairs $(K,C_K)$ where $C_K$ is defined as the following sum of matrix products:
\begin{equation}
C_K=\sum\limits_{\begin{footnotesize} \begin{array}{rr}I \in \Ind_A,\\ K-I \in \Ind_B \end{array}\end{footnotesize} } A_I B_{K-I}
\end{equation}
where $C_K \neq 0$. Therefore holds $e_C \leq e_A+e_B$ and
$\Eff_C=\Eff_A \cdot \Eff_B$ can be computed in the time $O(e_A\cdot e_B)$.

\smallskip
Similarly holds for computing the sum $C(S)=A(S)+B(S)$. Elements
of $\Eff_C$ are pairs $(K,C_K)$ where values $C_K$ are defined by
\begin{equation}
C_K=\left\{%
\begin{array}{ll}
A_K, & A_K\neq 0, B_K=0 \\
B_K, & B_K\neq 0, A_K=0 \\
A_K+B_K, & A_K\neq 0, B_K\neq 0\\
\end{array}%
\right.
\end{equation}
and satisfy $C_K \neq 0$. As in the previous case we can conclude that $e_C \leq \max\{e_A,e_B\}$ and $\Eff_C$
can be computed in time $O(\max\{e_A,e_B\})$.

\smallskip
Index sets corresponding to addition and multiplication of sparse matrices are equal to:
$$
\Ind_{A+B}=\Ind_A \cup \Ind_B, \quad \Ind_{AB}=\Ind_A + \Ind_B
$$
In view of $(\ref{Eff})$, we compute $\Eff^*_A\!\!=\!\!\{(I,A^*_I) \mid (I,A_I) \!\in \!\Eff_A\}$ in time $O(e_A)$.

\smallskip
Usually, coefficient matrices $A_I$ in the polynomial representation (\ref{tri1}), i.e. in the sparse representation (\ref{cetiri1}) are sparse.
Using this fact we can significantly improve our
sparse structure $\Eff$ by using an appropriate structure for these constant coefficient matrices.

\begin{dfn} For the constant matrix $A=[a_{ij}] \in \mathbf{C}^{m \times n}$, denote the following sparse structure:
\begin{equation}
\Sp_A=\left\{ (i,j,a_{ij}) \mid a_{ij} \neq 0 \right\}
\end{equation}
Denote by $s_A=|\Sp_A|$ the size of the structure $\Sp_A$.
\end{dfn}

Similarly as in the case of $\Eff_A$, we can define elementary operations on these sparse structures:
$$
\aligned
\Sp_A\!+\!\Sp_B\!&=\!\{ (i,j,a_{ij}+b_{ij}) \mid (i,j,a_{ij}) \in \Sp_A \vee (i,j,b_{ij}) \in \Sp_B,\,  a_{ij}\!+\!b_{ij}\neq 0 \} \\
\Sp_A \cdot \Sp_B\!&=\!\{ (i,j,c_{ij}) \mid c_{ij}\!=\!\!\sum a_{ik}b_{kj} \neq 0, (i,k,a_{ik})\! \in \!\Sp_A \wedge (k,j,b_{kj})\! \in \! \Sp_B \} \\
\Sp^*_A\!&=\!\{(j,i,a^*_{ij}) \mid (i,j,a_{ij}) \in \Sp_A\}
\endaligned
$$
In this way, we have the following improvement of the structure $\Eff$:
\begin{eqnarray}
\label{cetiri7}
\Eff'_A\!\!\!&=&\!\!\!\left\{(J,\Sp_{A_J})\,|\,A_J \neq 0,\ 0 \leq J \leq \deg A(S)\right\}\\
\!\!\!&=&\!\!\!\left\{\left( J,\{i,j,(A_J)_{ij}| (A_J)_{ij}\neq 0\}\right)\, |\,A_J \neq 0,\ 0 \leq J \leq \deg A(S)\right\}.\nonumber
\end{eqnarray}
It can be seen that the complexity of computing $\Sp_A+\Sp_B$ is
$O(s_A+s_B)$ and for $\Sp^*_A$ is $O(s_A)$. In the case of
multiplication the complexity depends on concrete implementation.
Suppose that $A \in \mathbf{C}^{m \times n}$ and $B \in \mathbf{C}^{n \times p}$.
If the triples are sorted lexicographically in $\Sp_A$ and in $\Sp_B$ then for every
$(i,k,A_{ik}) \in \Sp_A$ we need to find all $(k,j,B_{kj}) \in \Sp_B$,
i.e. all triples in $\Sp_B$ which begin by $k$. If we denote this number by $s^{(k)}_B$:
\begin{equation*}
s^{(k)}_B=|\{ (k,j,b_{kj}) \in \Sp_B \mid j=1,\ldots,p\}|
\end{equation*}
then the complexity of multiplication $\Sp_A\cdot \Sp_B$ is:
\begin{equation}
O\left(\sum_{(i,k,a_{ik}) \in \Sp_A} s^{(k)}_B+m \cdot p\right).
\end{equation}
The last addend in (4.8) comes from the fact that we
need to construct the sparse structure $\Sp_C$ for the matrix $C=AB \in \mathbf{C}^{m \times p}$.

\smallskip
We implemented the sparse structure $\Sp$ in {\ssr MATHEMATICA} as the structure {\tt SparseArray}.
Mathematica offers a sparse representation for matrices, vectors, and tensors with {\tt SparseArray} \cite{Wol1}, \cite{Wol}.
Both of the expressions

$\text{\tt SparseArray}[\{i_1,j_1\}-\!\!\!>v_1,\{i_2,j_2\}-\!\!\!>v_2,\ldots ,]$

$\text{\tt SparseArray}[\{\{i_1,j_1\},\{i_2,j_2\},\ldots\}-\!\!\!>\{v_1,v_2,\ldots \}]$

\noindent represent the sparse array with elements in positions $\{i_k,j_k\}$ having values $v_k$.

\noindent Operations on sparse matrices are all equivalent to the operations on dense matrices  \cite{Wol1}, \cite{Wol}:
{\tt Plus}({\tt +}) for matrix addition, {\tt Dot} ({\tt .}) for matrix multiplication, {\tt Times} ({\tt *}) for multiplication by scalar, etc.

\smallskip
Therefore, in our implementation we have
$$
\Eff'_A=\left\{(J,\, \mathrm {\tt SparseArray}[A_J])\,|\,A_J \neq 0,\ 0 \leq J \leq \deg A(S)\right\}.
$$
Shown fact that basic operations are the same for dense and sparse matrices
allows us to use the same procedures for basic operations on $\Eff$ in cases when $\Sp$ is embedded in $\Eff$ and when it is not.
In procedural programming languages we can decide to use $\Sp$ or not in the beginning of algorithm,
depending of the structure of input matrices $A(S)$, $M(S)$ and $N(S)$.
Similarly, it is possible to change the choice of one between these two variants of the structure $\Eff$ during the algorithm implementation.

\medskip
In the second type of the sparse structure for polynomial matrices
we represent the matrix $A(S)$ in the form $A(S)=[a_{ij}(S)]$, where $a_{ij}(S)$ are scalar polynomials,
and construct effective sparse structures $\Eff_{a_{ij}}$ for each $a_{ij}(S)$.
Effective structure $\Eff_a$ for the scalar polynomial $a(S)=\sum_{I=1}^{\deg a(S)} a_IS^I$
is defined similarly as in the matrix case $(\ref{cetiri1})$:
$$\Eff_a=\{ (J,a_J) \mid a_J \neq 0, \ 1 \leq J \leq \deg a(S) \}.$$
Such sparse representation we denote by $\Ef$, and have $\Ef_{A}=[\Eff_{a_{ij}}]$.
If we use notations $\ef_A=\sum_{i=1}^m\sum_{j=1}^n e_{a_{ij}}$, then the complexity for the addition is
$$O\left(\sum_{i=1}^m\sum_{j=1}^n e_{a_{ij}}+e_{b_{ij}}\right)=O(\ef_A+\ef_B).$$
After the notations $\row(B,k)=\sum_{j=1}^p e_{b_{kj}}$ and $\col(A,k)=\sum_{i=1}^m e_{a_{ik}}$ we
conclude that the complexity of the matrix multiplication $C=AB$ is equal to
\begin{eqnarray*}
O\left(\sum_{i=1}^m\sum_{k=1}^n\sum_{j=1}^p e_{a_{ik}}e_{b_{kj}}\right)\!\!\!\!&=&\!\!\!\!O\left(\sum_{k=1}^n\sum_{i=1}^me_{a_{ik}}\row(A,k)\right)\\
&=&O\!\!\left(\sum_{k=1}^n \col(A,k)\row(B,k)\right)
\end{eqnarray*}
for the multiplication.

\smallskip
Polynomials in {\ssr MATHEMATICA} are represented in the internal form using the little modified $\Ef$
sparse structure. For example, two-variable polynomial $p(s_1,s_2)=4 s_1^9 s_2^{10}+s_2^3+s_1^2 s_2^2+3 s_1^3 s_2+s_2+2 s_1^2+3 s_1+10$
is represented in the following {\ssr MATHEMATICA} internal form:
\begin{center}
\begin{verbatim}
Plus[
10,
Times[3, s1],
Times[2, Power[s1, 2]],
s2,
Times[3, Power[s1, 3], s2],
Times[Power[s1, 2], Power[s2, 2]],
Power[s2, 3],
Times[4, Power[s1, 9], Power[s2, 10]]
].
\end{verbatim}
\end{center}
The last expression is obtained by using {\ssr MATHEMATICA} function {\tt FullForm[E]} which
returns an internal representation of the expression $E$ \cite{Wol1}, \cite{Wol}.
This internal form of the polynomial $p(S)$, at the top level is the list with length $\ef_p$ with the head {\tt Plus}.
Each element of this list contains the exponent $J=(j_1,j_2)$ and the value $p_J$ (values $j_1=0,1$ and $j_2=0,1$ and are not shown),
hence the length of each element is $O(1)$. Also the size of whole structure is $O(\ef_{p(s)})$.
Therefore, we can use this natural polynomial representation in {\ssr MATHEMATICA} and built-in elementary operators
to implement the effective partitioning method using $\Ef$ structure. The complexity of these
built-in operations are the same as corresponding operations defined for $\Ef$ structure.

\smallskip
The next algorithm is the effective partitioning method for computing
the we\-igh\-ted Moore-Penrose inverse of polynomial matrices, suitable for sparse matrices.
Generally, the same method can be used with both two
presented sparse structures. Therefore, we will denote {\it general sparse structure} with $\e$, which can be exchanged
either by $\Eff$ or $\Ef$. Also by $\O$ we will denote the general effective structure of an appropriate zero matrix.
We will use the same symbol for the effective structure of the number $0$.

\begin{alg} $($Computing the weighted Moore-Penrose inverse $A(S)^{\dagger }_{M(S),N(S)}$ of sparse matrix $A(S))$.

Input: Effective structures of matrices $A(S)$, $M(S)$, $N(S)$.

\begin{itemize}
\item[{\it Step 1.\/}] In the case $\e_{a_1} \neq \O$ compute initial values:
$$
\e_{Z_1}=\e^*_{a_1} \cdot \e_M, \quad \e_{Y_1}=\e^*_{a_1}\cdot \e_M\cdot \e_{a_1}.
$$
If $\e_{a_1}=\O$, then set $\e_{Z_1}=\O$ and $\e_{Y_1}=\e_1$, where $\e_1$ is the corresponding
sparse structure of the number $1$.

\smallskip
\item[{\it Step 2.\/}] Recursive step: For $i=2,\ldots,n$ perform the following steps
\begin{itemize}
\item[{\it Step 2.1\/}] Compute: $\e_{d_i}=\e_{Z_{i-1}}\cdot \e_{a_i}$
\item[{\it Step 2.2\/}] Compute: $\e_{c_i}=\e_{a_i}\cdot \e_{Y_{i-1}}-\e_{A_{i-1}} \cdot \e_{d_i}$
\item[{\it Step 2.3\/}] If $\e_{c_i} \neq \O$ then compute $\e_{V_i}$ and $\e_{W_i}$ using
$$
\aligned
\e_{V_i}&=\e_{Y_{i-1}} \cdot \e^*_{c_i} \cdot \e_M \\
\e_{W_i}&=\e^*_{c_i} \cdot \e_M \cdot \e_{c_i}.
\endaligned
$$
Otherwise use the following formulae:
$$
\aligned
\e_{V_i}&=\e_{\oo{\Delta}_i} \cdot (\e^*_{d_i} \cdot \e_{N_{i-1}}-\e^*_{l_i} \cdot \e_{Y_{i-1}})\cdot \e_{Z_{i-1}} \\
\e_{W_i}&=\e_{\ov{\Delta}_i} \cdot \e^*_{Y_{i-1}} \cdot \e_{Y_{i-1}},
\endaligned
$$
where the structures $\oo{\Delta}_i$ and $\ov{\Delta}_i$ are defined by:
$$
\aligned
\e_{\oo{\Delta}_i}\!\!&=\e_{\psi_i} \cdot \e^*_{Y_{i-1}} \cdot \e_{Y_{i-1}}\\
\e_{\ov{\Delta}_i}\!\!&=(\e_{Y_{i-1}} \cdot (\e_{n_{ii}}\e_{Y_{i-1}}-\e_{d^*_i}\cdot \e_{l_i}-\e_{l^*_i}\cdot \e_{d_i})+
               \e_{d^*}\cdot \e_{N_{i-1}} \cdot \e_{D_i})\cdot \e_{\oo{N}_{i-1}}\\
               &\ -\e_{Y_{i-1}}\cdot \e_{l^*_i}\cdot \e_{\varphi_i}
\endaligned
$$

We used sparse representations for temporary variables $\varphi_i$ and $\psi_i$, defined in $(\ref{fipsi})$:
$$
\aligned
\e_{\varphi_i}&=(\e_{Y_{i-1}}\cdot \e_{I} -\e_{Z_{i-1}} \cdot \e_{A_{i-1}})\cdot \e_{\ov{N}_{i-1}} \cdot \e_{l_i} \\
\e_{\psi_i}&=\e_{Y_{i-1}}\cdot \e_{\oo{N}_{i-1}}.\\
\endaligned
$$

\item[{\it Step 2.4.\/}] Now compute $\e_{Z_i}$ and $\e_{Y_i}$ using:
$$
Z_i=\bmatrix \Theta_i \\ \Psi_i \endbmatrix, \qquad
\e_{Y_i}=\e_{\psi_i}\cdot \e_{W_i}\\
$$
Structures $\e_{\Theta_i}$ and $\e_{\Psi_i}$ are defined by:
$$
\aligned
\e_{\Theta_i}&=\e_{Z_{i-1}} \cdot \e_{\oo{N}_{i-1}} \cdot \e_{W_i}-\e_{d_i}\cdot
\e_{\oo{N}_{i-1}}\cdot \e_{V_i}-\e_{\varphi_i}\e_{V_i}\\
\e_{\Psi_i}&=\e_{\psi_i}\cdot \e_{V_i}\\
\endaligned
$$

If we use $\Ef$ or $\Eff$ sparse structure, $\e_{Z_i}$ is equal respectively to:
\begin{equation}
\aligned
\Ef_{Z_i}=&\bmatrix \Ef_{\Theta_i} \\ \Ef_{\Psi_i} \endbmatrix \label{joineff} \\
\Eff_{Z_i}\!=&\!
           \left\{ \! \left( j, \bmatrix (\Theta_i)_j \\ (\Psi_i)_j \endbmatrix \right) \mid (j,(\Theta_i)_j)\! \in \!\Eff_{\Theta_i}, (j,(\Psi_i)_j)
             \! \in \!\Eff_{\Psi_i}\! \right\}\\
&\cup \left\{\left( j, \bmatrix (\Theta_i)_j \\ 0 \endbmatrix \right) \mid (j,(\Theta_i)_j) \in \Eff_{\Theta_i}, (\Psi_i)_j=0 \right\}\\
&\cup \left\{\left( j, \bmatrix 0 \\ (\Psi_i)_j \endbmatrix
\right) \mid (\Theta_i)_j=0, (j,(\Psi_i)_j) \in \Eff_{\Psi_i}
\right\}
\endaligned
\end{equation}

\item[{\it Step 2.5.\/}] Find the polynomials $Z_i(S)$ and $Y_i(S)$ from its effective structures and compute:
\begin{equation}
X_{i}(S) = \frac{Z_i(S)}{Y_i(S)},
\end{equation}
Cancel the common multipliers in numerator $Z_i(S)$ and denominator $Y_i(S)$,
recompute (if necessary) effective structures and continue with the next $i$.
\end{itemize}
\item[{\it Step 3.\/}] The stopping criterion is $i=n$. In this case is $A_{M(S),N(S)}^\dagger(S) =X_n(S)$.
\end{itemize}

\end{alg}

Similarly we can derive a modification of the method introduced in Theorem 3.2 for
computing the inverse matrix $N^{-1}_i(S)$ in the polynomial form:

\begin{equation}
N^{-1}_i(S)=\frac{\ov{N}_i(S)}{\oo{N}_i(S)}
\end{equation}

\noindent
\begin{alg} $($Effective computation of $N^{-1}_i(S)$, for $i=1,\ldots,n)$.

\smallskip
Input: Effective structure of positive definite Hermitian polynomial matrix $N(S)$ of the order $n$.
Notations are the same as in Theorem 3.2.

\begin{itemize}
\item[{\it Step 1.\/}] Generate initial values: $\ov{N}_1=I$ and $\oo{N}_1=n_{11}$ and corresponding effective structures.
\item[{\it Step 2.\/}] Recursive step: For $i=2,\ldots,n$ perform following steps:
\begin{itemize}

\item[{\it Step 2.1.\/}] Compute: $\e_{\oo{H}_i}=\e_{n_{ii}} \cdot \e_{\oo{N}_{i-1}} - \e^*_{l_i} \cdot \e_{\ov{N}_{i-1}} \cdot \e_{l_i}$.

\smallskip
\item[{\it Step 2.2.\/}] Compute: $\e_{\ov{F}_i}=-\e_{\ov{N}_{i-1}}\cdot \e_{l_i}$.

\smallskip
\item[{\it Step 2.3.\/}] Compute: $\e_{\ov{E}_i}=\e_{\ov{N}_{i-1}}-\e_{F_i} \cdot\e^*_{F_i}$.

\smallskip
\item[{\it Step 2.4.\/}] Generate:
$$
\aligned
\ov{N}_i(S)&=
\bmatrix
\ov{E}_{i-1}(S) & \oo{N}_{i-1}(S) \cdot \ov{F}_i(S) \\
\oo{N}_{i-1}(S) \cdot \ov{F}^*_i(S) & \oo{N}_{i-1}^2(S)
\endbmatrix\\
\oo{N}_i(S)&=\oo{N}_{i-1}(S)\cdot\oo{H}_i(S)
\endaligned
$$

As in the previous algorithm, we have also two different representations for $\Ef$ and $\Eff$ sparse structures.
These relations are similar to $(\ref{joineff})$.

\end{itemize}
\item[{\it Step 3.\/}] Stop criterion for $i=n$. Inverse matrix $N_k^{-1}(S)$, for every $k=1,\ldots,n$ is equal to:
\begin{eqnarray}
N_{k}^{-1}(S)=\frac{\ov{N}_k(S)}{\oo{N}_k(S)}
\end{eqnarray}
\end{itemize}

\end{alg}

\section{Examples}\setcounter{equation}{0}

We implemented algorithms 2.1, 2.2, 4.1 and 4.2 in the programming language {\ssr MATHEMATICA}.
An implementation of the $\Eff$ sparse structure is also made. Functions {\tt WPolyEf} and
{\tt WPolyEff} implement Algorithm 4.1 using respectively $\Ef$ and $\Eff$ sparse strucure.
All basic operations for $\Eff$ sparse structure (functions {\tt Add}, {\tt Sub}, {\tt Muls}, {\tt Mul} and {\tt TE} corresponding
to the addition, subtraction, multiplication by scalar, multiplication and conjugate-transposion respectively) are also implemented.

\begin{exm} Let us find the weighted Moore-Penrose inverse of the following two-variable polynomial matrix $A(x,y)$:
$$
A(x,y)=\bmatrix
1 - 3 x & 5 + 9 x - 10 y & 16 + 8 x + 2 y \\
-7 + 9 x - 8 y& 8 + 5 x - y & 4 + 2 x + 3 y \\
7 - x - 8 y & 16 - 2 x - 6 y & -3 - 2 x - 4 y
\endbmatrix
$$
with respect to the following matrices $M(x,y)$ and $N(x,y)$:
\begin{eqnarray*}
M(x,y)\!\!&=&\!\!
\bmatrix
-20 - x - \conjugate{x}& -8 - 7 x -4 \conjugate{x} & -2 (8 + 3 x + 4 \conjugate{x}) \\
-8 - 4 x - 7 \conjugate{x}& -20 + 7 x + 7 \conjugate{x}&
2 (5 x - \conjugate{x})\\
-2 (8 + 4 x + 3 \conjugate{x})& -2 (x -5 \conjugate{x})& 7 (-2 + x + \conjugate{x})\\
\endbmatrix
\\
{} \\
N(x,y)\!\!&=&\!\!
\bmatrix
16 + 7 x + 7 \conjugate{x}& 7 - 6 x - 2 \conjugate{x}& 6 - 10 x - 3 \conjugate{x}\\
7 - 2 x - 6 \conjugate{x}& -2 (3 + 5 x + 5 \conjugate{x})& -2 (6 + 4 x + 3 \conjugate{x})\\
6 - 3 x - 10 \conjugate{x}& -2 (6 + 3 x + 4 \conjugate{x})& -3 (-6 + x + \conjugate{x})
\endbmatrix .
\end{eqnarray*}

The obtained weighted Moore-Penrose inverse is:
\begin{small}
$$
\aligned &X(x,y)=A^\dagger_{M(x,y),N(x,y)}(x,y)=\left( 60 x^3\!-\!5 y
x^2\!-\!540 x^2\!+\!51 y x\!+\!779 x\!-\!42 y\!-\!435\right)\!^{-1} \\
&\times \bmatrix
-5 x^2+51 x-42 & -3 x^2+8 x-13 & -3 x^2+33 x-4 \\
-30 x^2+71 x+15 & 42 x^2-5 y x-33 x+y+15 & -18 x^2-63
x+10 y+105 \\
-2 \left(\!10 x^2\!-\!19 x\!+\!12\!\right) & 2 \left(\!18 x^2\!-\!2y x\!-\!29 x\!+\!2 y\!+\!17\right)
                   & -24 x^2\!+\!y x\!+\!42 x\!-\!2 y\!-\!23
\endbmatrix
\endaligned
$$
\end{small}
\end{exm}

Let us notice that degrees of intermediate results in algorithms 4.1 and 4.2 are much greater than the degrees of $A, M, N$ and $X$ (maximum
degree in this example are $874$ and $122$ of the variables $x$ and $y$ respectively). This is the reason why the algorithms
for computing the weighted Moore-Penrose inverse for polynomial matrices are very slow (working time of the function {\tt WPolyEff}
for last example is $172.922$ seconds). As we will see in the sequel, when matrices $A$, $M$ and $N$ are sparse, corresponding intermediate results are
also sparse. Therefore, sparse structures introduced in the previous section improve the working time of the implementation.

\smallskip
Algorithm 4.1 is tested on several random generated test examples.
We tested variants of algorithm 4.1 using $\Ef$ and $\Eff$ sparse
structures separately. In this test, matrices $A(S)$, $M(S)$ and
$N(S)$ were complex polynomial matrices of one variable $s$ (i.e. holds $S=(s,\overline{s})$).

\smallskip
We made testing for two different classes of matrices: sparse and dense.
The measures representing sparsity of a given polynomial matrix
are the same as in \cite{Marko} (definitions 6.1 and 6.2). We are
now restating these two definitions and generalizing them to the multi-variable complex polynomial matrices.

\begin{dfn} For a given matrix $A(S)=[a_{ij}(S)] \in \mathbf{C}[S]^{m \times n}$ (polynomial or constant), {\bf the first sparse number} $sp_1(A)$
is the ratio of the total number of non-zero elements and total number of elements in $A(S)$:
$$sp_1(A(S))=\frac{\left| \{(i,j) \,|\, a_{ij}(S)\neq 0\} \right|}{m \cdot n}.$$
\end{dfn}

The first sparse number represents the density of non-zero elements and it is between $0$ and $1$.

\begin{dfn} For a given polynomial matrix $A(S) \in \mathcal \mathbf{C}[S]^{m \times n}$ and $S=(s_1,\ldots,s_p)$, {\bf the second sparse
number} $sp_2(A(S))$ is the following ratio:
$$sp_2(A(S))\!\!=\!\!\frac{\#\{(i,j,k_1,\ldots,k_p) \mid 0\! \leq \!k_j\! \leq\! {\rm deg}_{s_j}\!A(S),
   {\rm Coef}(a_{ij}(S),s_1^{k_1}\cdots s_p^{k_p}\!)\!\neq \!0\!\}}
{{\rm deg}_{s_1}A\cdots{\rm deg}_{s_p}A\cdot m \cdot n}.$$
By ${\rm Coef}(P(S),s_1^{k_1}\cdots s_p^{k_p})$ we denoted
the coefficient corresponding to $s_1^{k_1}\cdots s_p^{k_p}$ in polynomial $P(S)$.
\end{dfn}

The second sparse number represents density of non-zero coefficients contained
in elements $a_{ij}(S)$, and it is also between $0$ and $1$.

Results are presented in the next table (column $d$ states for the
degree of corresponding matrix polynomials $A(S)$, $M(S)$ and $N(S)$):

\begin{center}
\begin{tabular}{|c|c|c|c|c|}
\hline $m$ & $n$ & $d$ & Alg 4.1 & Alg. 4.1 \\
&&& with $\Ef$ & with $\Eff$\\
\hline 2 & 2 & 1 & 0.14 & 0.188\\
\hline 2 & 2 & 2 & 0.65 & 1.24\\
\hline 2 & 2 & 3 & 1.92 & 3.93\\
\hline 3 & 3 & 1 & 1.34 & 1.32\\
\hline 3 & 3 & 2 & 9.01 & 11.81\\
\hline 3 & 3 & 3 & 34.39 & 48.13\\
\hline 4 & 4 & 1 & 7.87 & 6.74\\
\hline 4 & 4 & 2 & 69.31 & 64.48\\
\hline 4 & 4 & 3 & 461.07 & 594.98\\
\hline 5 & 5 & 1 & 49.13 & 58.48\\
\hline 5 & 5 & 2 &  309.38 & 330.32\\
\hline
\end{tabular}
\hskip 1truecm
\begin{tabular}{|c|c|c|c|c|}
\hline $m$ & $n$ & $d$ & Alg 4.1 & Alg. 4.1 \\
&&& with $\Ef$ & with $\Eff$\\
\hline 2 & 2 & 1 & 0.06 & 0.89\\
\hline 2 & 2 & 2 & 0.25 & 0.46\\
\hline 2 & 2 & 3 & 0.60 & 1.23\\
\hline 3 & 3 & 1 & 0.47 & 0.68\\
\hline 3 & 3 & 2 & 4.60 & 7.18\\
\hline 3 & 3 & 3 & 14.89 & 24.65 \\
\hline 4 & 4 & 1 & 6.10 & 6.18\\
\hline 4 & 4 & 2 & 34.95 & 39.68\\
\hline 4 & 4 & 3 & 256.31 & 299.61\\
\hline 5 & 5 & 1 & 30.85 & 39.43\\
\hline 5 & 5 & 2 & 246.32 & 283.12\\
\hline
\end{tabular}
$sp_1(A(S))=0.9, \ sp_2(A(S))=0.9$ \hskip 1truecm
$sp_1(A(S))=0.7, \ sp_2(A(S))=0.5$
\vskip0.6truecm

\begin{tabular}{|c|c|c|c|c|}
\hline $m$ & $n$ & $d$ & Alg 4.1 & Alg. 4.1 \\
&&& with $\Ef$ & with $\Eff$\\
\hline 2 & 2 & 1 & 0.04 & 0.112\\
\hline 2 & 2 & 2 & 0.11 & 0.263\\
\hline 2 & 2 & 3 & 0.422 & 1.303\\
\hline 3 & 3 & 1 & 0.281 & 0.972\\
\hline 3 & 3 & 2 & 1.367 & 3.505\\
\hline 3 & 3 & 3 & 5.808 & 18.449\\
\hline 4 & 4 & 1 & 1.613 & 5.549\\
\hline 4 & 4 & 2 & 12.134 & 27.113\\
\hline 4 & 4 & 3 & 55.139 & 107.27\\
\hline 5 & 5 & 1 & 7.475 & 13.582\\
\hline 5 & 5 & 2 & 84.712 & 139.681\\
\hline
\end{tabular}
\hskip 1truecm
\begin{tabular}{|c|c|c|c|c|}
\hline $m$ & $n$ & $d$ & Alg 4.1 & Alg. 4.1 \\
&&& with $\Ef$ & with $\Eff$\\
\hline 2 & 2 & 1 & 0.032 & 0.105\\
\hline 2 & 2 & 2 & 0.069 & 0.190\\
\hline 2 & 2 & 3 & 0.187 & 0.713\\
\hline 3 & 3 & 1 & 0.185 & 0.675\\
\hline 3 & 3 & 2 & 0.628 & 2.944\\
\hline 3 & 3 & 3 & 1.031 & 3.275 \\
\hline 4 & 4 & 1 & 0.987 & 4.344\\
\hline 4 & 4 & 2 & 6.087 & 25.263\\
\hline 4 & 4 & 3 & 27.466 & 176.581 \\
\hline 5 & 5 & 1 & 3.294 & 15.853 \\
\hline 5 & 5 & 2 & 42.159 & 171.416\\
\hline
\end{tabular}
$sp_1(A(S))=1, \ sp_2(A(S))=0.2$ \hskip 1truecm
$sp_1(A(S))=0.2, \ sp_2(A(S))=0.2$
\end{center}

All presented processor times are in seconds and the sparse numbers for matrices $M(S)$ and
$N(S)$ are the same as corresponding sparse numbers for $A(S)$. Every processor time
is obtained by averaging working times of 15 different randomly generated test cases.
Testing was done on Intel Pentium 4 processor at 2.6GHz and {\ssr MATHEMATICA} 5.2. We can
notice that Algorithm 4.1 with an $\Ef$ structure showed best
timings on all test cases. We have already mentioned that an
$\Ef$ sparse structure is already implemented in {\ssr MATHEMATICA}.
In the implementation we used standard built-in operators for
manipulation with matrices in $\Ef$ structure.

\smallskip
The first table (when $sp_1(A(S))=sp_2(A(S))=0.9$) corresponds to dense matrices. In this case, sparse
structures are not so effective because there are a lot of non-zero elements in all matrices and non-zero
coefficients in polynomials. But we can notice significant improvement in working time when is applied
$\Ef$ structure against the case when $\Eff$ structure is applied. This difference mainly comes from the fact
that $\Ef$ structure is implemented by {\ssr MATHEMATICA} built-in operations.

\smallskip
The second case (when $sp_1(A(S))=0.7$ and $sp_2(A(S))=0.5$) represents sparse matrices. We can notice that
working times are significantly less than in the first case. Also here $\Ef$ structure produces less working
times than $\Eff$.

\smallskip
In the third and fourth case (when $sp_1(A(S))=1$ and $sp_2(A(S))=0.2$, and $sp_1(A(S))=sp_2(A(S))=0.2$, respectively) we deal with
matrices whose entries are very sparse polynomials.
Moreover, in the fourth case we work with matrices with only few non-zero elements.
In the fourth case, smallest average working times are obtained for all considered
matrix dimensions and degrees. Also we can notice that as sparse numbers decrease, the average working times also decrease (for constant
matrix dimensions and degree). This holds for both sparse structures and verifies the theoretical results about sparse structures $\Ef$ and $\Eff$
in practice.

\smallskip
We also considered simpler case: when all
input matrices ($A(S)$, $M(S)$ and $N(S)$) and variables
$s_1,\ldots,s_p$ are assumed to be real. In that case we have only
$p$ variables and conjugate-transpose operation reduces only to
transpose. We also should suppose that matrices $M(S)$ and $N(S)$
are symmetric in that sense. Algorithms 4.1 and 4.2 remains the same
except we should change the definition of conjugate-transpose
operator (also the implementations in {\ssr MATHEMATICA}). This case
is considered in \cite{TSP} and algorithms 4.1 and 4.2 are an
effective versions of corresponding algorithms 3.1 and 3.2 in \cite{TSP}.
Here working times of the algorithms are significantly less, and also the inverses has much smaller degrees.
Results obtained in this special case are presented in the following table:

\smallskip
\begin{center}
\begin{tabular}{|c|c|c|c|c|c|c|}
\hline $m$ & $n$ & $d$ & Alg 2.1 & Alg 4.1 & Alg. 4.1 & Alg 3.1\\
&&&& with $\Eff$ & with $\Ef$ & from \cite{TSP} \\
\hline 3 & 3 & 1 & 0.32 & 0.23 & 0.10 & 0.94 \\
\hline 3 & 3 & 2 & 0.69 & 0.57 & 0.20 & 1.32 \\
\hline 3 & 3 & 3 & 0.82 & 1.17 & 0.43 & 1.84 \\
\hline 3 & 3 & 4 & 1.19 & 2.15 & 0.73 & 2.38\\
\hline 4 & 3 & 1 & 0.76 & 1.26 & 0.14 & 1.29\\
\hline 4 & 3 & 2 & 1.29 & 0.65 & 0.31& 2.12\\
\hline 4 & 3 & 3 & 2.14 & 1.32 & 0.59 & 2.42 \\
\hline 4 & 3 & 4 & 2.84 & 2.26 & 1.01 & 2.93\\
\hline 5 & 5 & 1 & 3.48 & 1.45 & 1.01 & 3.56\\
\hline 5 & 5 & 2 & 5.90 & 4.54 & 2.92 & 4.92\\
\hline 5 & 5 & 3 & 9.18 & 8.79 & 6.82 & 8.27\\
\hline 5 & 5 & 4 & 12.15 & 15.87 & 10.85 & 10.34\\
\hline 6 & 6 & 1 & 7.98 & 2.65 & 2.17 & 8.16\\
\hline 6 & 6 & 2 & 12.93 & 8.20 & 7.31 & 11.32\\
\hline 6 & 6 & 3 & 21.76 & 18.29 & 13.53 & 19.42\\
\hline
\end{tabular}\\
$sp_1(A(S))=0.7$, $sp_2(A(S))=0.7$
\end{center}

\smallskip
It can be seen from the table that here in all cases $\Ef$ structure was better than $\Eff$ (both with using Algorithm 4.1).
Both effective algorithms was significantly better than Algorithm 2.1 (for rational matrices) and Algorithm 3.1 from \cite{TSP}.
For smaller values of $d$, Algorithm 2.1 was better than Algorithm 3.1 from \cite{TSP} due to the implementation details.

\smallskip
All presented results leads us to the same conclusion:
the best choice for computing weighted Moore-Penrose inverse for polynomial matrices is Algorithm 4.1 with the sparse structure $\Ef$.

\section{Conclusion}
We extend the algorithm for computing the weighted Moore-Penrose from
\cite{Wang1} to the set of multiple-variable rational matrices with complex coefficients.
We adapt previous algorithm to the set of polynomial matrices.
We consider two effective structures which make use of only nonzero addends in
polynomial matrices and improve previous results on the set of sparse matrices.
In the last section we presented an illustrative example and compared various algorithms.

\enddocument
\bye